\documentclass{article}
\usepackage{graphicx} 
\usepackage{amsmath,amssymb,amsthm,amsfonts,mathptmx}

\usepackage{authblk}

\usepackage{xcolor}     
\usepackage{hyperref}   
\usepackage{tabularx}
\usepackage{url}        
\usepackage{multirow}   

\title{Downtime Required for Bitcoin Quantum-Safety}
\author{Jamie J. Pont} 
\author{Joseph J. Kearney} \author{Jack Moyler} 
\author{Carlos A. Perez-Delgado\footnote{c.perez@kent.ac.uk}}
\affil{School of Computing, University of Kent,\\ Canterbury, Kent, CT2 7NF, United Kingdom}
\date{}

\begin{document}

\newcommand{\citethis}[1]{\color{blue}\textbf{(CITATION NEEDED: #1)}\color{black}}
\newcommand{\todo}[1]{\color{orange}\textbf{(TO DO: #1)}\color{black}}
\newcommand{\joseph}[1]{\color{red}\textbf{(JOSEPH: #1)}\color{black}}
\newcommand{\jamie}[1]{\color{purple}\textbf{(JAMIE: #1)}\color{black}}
\newcommand{\jack}[1]{\color{green}\textbf{(JACK: #1)}\color{black}}
\newcommand{\carlos}[1]{\color{teal}\textbf{(CARLOS: #1)}\color{black}}

\maketitle

\begin{abstract}

Quantum devices capable of breaking the  public-key cryptosystems that Bitcoin relies on to secure its transactions are expected with reasonable probability within a decade.
Quantum attacks would put at risk the entire Bitcoin network---which has  an estimated value of around 500 billion USD. To prevent this threat, a proactive approach is critical. The only known way to prevent any such attack is to upgrade---\emph{replace}---the currently used public-key cryptosystems, namely ECDSA, with so-called post-quantum cryptosystems which have no known vulnerabilities to quantum attacks. In this paper, we analyse the technical cost of such an upgrade. We calculate a non-tight lower bound on the cumulative downtime required for the above transition to be 1827.96 hours, or 76.16 days. We also demonstrate that the transition needs to be fully completed \emph{before} the availability of ECDSA-256 breaking quantum devices---in order to ensure Bitcoin's ongoing security. The conclusion is that the Bitcoin upgrade to quantum-safe protocols needs to be started as soon as possible in order to guarantee its ongoing operations.
\end{abstract}

\section{Introduction}

The advent of quantum technologies represents a clear and present threat to the cybersecurity of all network systems\cite{mosca2018cybersecurity}. In short, quantum computers are able to solve numerical problems---such as factoring and the discrete logarithm problem---which are intractable on classical computers. This makes all cryptosystems that rely on the difficulty of these problems---namely public-key ones such as RSA and Elliptic-Curve Cryptography (ECC)---unsafe, and therefore obsolete, once scalable universal quantum computers are made available. According to a survey of major quantum computation experts, the introduction of such devices can be expected, with roughly $31\%$ probability, within the next decade\cite{mosca2023quantum}.

This threat is particularly acute in cryptocurrencies and other blockchain-based technologies. This is because the service, and the cryptosystem, are intrinsically intertwined. As an example, a bank whose online services system is made vulnerable due to an unsecure cryptosystem, can bring such online services down, until a patch is applied. While this is costly, and disruptive, the bank would be able to continue to do its business in other ways: telephone banking and in-person banking. If a cryptocurrency's cryptographic protocols are made insecure---whether through the discovery of a heretofore unknown vulnerability to classical attackers, or through the development of quantum-enabled attackers---then the cryptocurrency completely ceases to exist as such.

Current major blockchains, including Bitcoin, exhibit major vulnerabilities to quantum attacks due, in part, to their use of public-key cryptosystems based on factoring and/or discrete-logarithm\cite{aggarwal2017quantum,kearney2021vulnerability,Bard2022}.

The solution to the above threat---both for the general cryptosystem case, and in the particular case of cryptocurrencies/blockchains---is widely understood to be the replacement of current-day public-key cryptosystems with what is known as \emph{``post-quantum''} cryptosystems. These are public-key cryptosystems whose security is based on the difficulty of numerical problems that are not known to be tractable on quantum devices\cite{bernstein2017post}.

Proverbially, this is easier said than done. Any \emph{upgrade} to a current-day cryptosystem to make it quantum-secure comes with an all-but inevitable downtime. 

Downtime should not be understood to be merely the situation when an online service is purposely taken down, \emph{i.e.} to apply a patch. Downtime occurs whenever a service is made unavailable, for whatever reason. These can include purposefully bringing the service offline, network outages, denial-of-service attacks, among other reasons.

Downtime can have a massive financial impact on any online service. It has been estimated that IT infrastructure downtime can have a cost anywhere between \$300,000 per hour and \$1,000,000 per hour\cite{gartner2016,idc2014}. Given that Bitcoin sees a throughput of approximately 1.45 billion USD per hour\cite{blockchain_btc}, the impact of downtime to a blockchain network could be even more pronounced.

Making Bitcoin quantum-safe will, at the very least, require a two-pronged approach. First is the adoption of a rule requiring all newly created UTXOs (\textit{unspent transaction outputs}---a record of unspent Bitcoin units; in other words, the Bitcoin equivalent of a ``bank account'') to be quantum-safe. UTXOs require, among other data, a private-public key-pair. Currently, this private-public key-pair uses ECDSA (\textit{Elliptic Curve Digital Signature Algorithm}), making it susceptible to quantum attacks. 

Second is to move all currently unspent Bitcoin from current-technology UTXOs to quantum-safe UTXOs. It is this second step that incurs a downtime cost. Bitcoin has a ten-minute \emph{blocktime;} meaning that a new block is added every ten minutes. Every block has a finite size, allowing only for so many transactions. This means that \emph{``upgrade''} transactions are competing with \emph{``regular''} transactions for the same limited space on each block. The larger percentage of the block is dedicated to upgrade transactions, the larger the degradation of the quality-of-service for regular transactions will be on a network that is already heavily criticised for its extremely long times to complete transactions\cite{ke2020bitcoin}. Conversely, the shorter the percentage of the block is dedicated to the upgrade, the longer the overall upgrade will take.

Stewart \emph{et al.} give a detailed analysis of the above approach\cite{stewart2018committing}, and the consensus seems to be, both among the scientific and Bitcoin communities, that this approach is both feasible and sufficient for quantum safety.

In this paper, we identify two major reasons for concern, in the absence of any further major changes to the protocol. First, is the overall downtime.  We calculate that an absolute lower bound on the  Bitcoin downtime required to move all unspent Bitcoin, as it stands today, to quantum-safe UTXOs is approximately 76 days. This is the total Bitcoin processing needed to upgrade all current UTXOs to quantum-safe private-public key-pairs, if the network is used \emph{for absolutely nothing else.} This downtime can  be amortised over a longer period of time. For instance, the process can be done over a period of approximately 152 days, assuming that 50\% network bandwidth is permitted. This would make only half of each block available for regular transactions, essentially doubling the transaction time during this period.

Clearly, both scenarios above are untenable. This may lead one to suggest amortising the upgrade process over the \emph{longest possible} time period, in order to minimise the network impact. This leads us to our second reason for concern. For reasons explained in the next section, many authors implicitly assume, and the work by Stewart \emph{et. al.}\cite{stewart2018committing} explicitly assumes, that not all current UTXOs need to be upgraded to quantum-safe protocols before scalable quantum computers are introduced. In the next section we show that this assumption is \emph{incorrect} by presenting what we call a \emph{Just-In-Tme} (JIT) quantum attack. The existence of such an attack means that, in order to ensure quantum safety, the procedure of moving \emph{all} unspent Bitcoin to quantum-safe UTXOs must be completed before quantum computers capable of breaking 256-bit ECDSA keys are developed.

Together, both of the above concerns make it clear that the current trajectory of Bitcoin is untenable. In order to guarantee quantum-safety, Bitcoin needs to greatly accelerate the timescale in which it begins, in earnest, its move to quantum-safe protocols; or it needs to explore more sophisticated approaches to ensure its quantum safety; or, more robustly, do both.

\section{JIT Quantum Attacks \& the Quantum-Safe Deadline}
\label{sec:jit}

In the previous section, we mentioned that the  work by Stewart \emph{et al.}\cite{stewart2018committing} explicitly assumes, and most authors assume either explicitly or implicitly, that not all current UTXOs need to be upgraded to quantum-safe protocols before scalable quantum computers are introduced.
The reason for this is how Bitcoin transactions currently work, and how they did in the past.

Every transaction in Bitcoin consists of inputs (payers) and outputs (payees). Previously, the agent behind every input (payer) and behind every output (payee) would have to make their public-key made publicly available for the transaction to go forward. This was changed in 2012\cite{BIP16} so that that transaction outputs use a \emph{hash} of the payee public-key. This means that since 2012, Bitcoin holders need only divulge their public-key when transacting out (\emph{i.e.} paying) Bitcoin. 

This leads to a natural conclusion that all UTXOs whose public-key has not been divulged are safe from quantum attacks; and hence need not be upgraded with any urgency.

However, consider the following scenario. Quantum technology has advanced to where an attacker, Charly, has access to a quantum computer capable of breaking ECDSA. Alice, whose (non-quantum safe) public-key  remains previously undivulged, attempts a transaction to a quantum-safe UTXO (either their own, to perform the aforementioned safety upgrade, or someone else's). 

In order to initiate the transaction, Alice necessarily divulges her public-key. The transaction goes into the transaction pool, to wait for a miner to pick up the transaction and add it to a mined block. At that moment, Charly downloads the publicly available public key and uses it to calculate the private-key using his quantum device. Charly then uses that private key to initiate a transaction of his own. He can then use various mechanisms---both widely available such as paying higher fees to encourage miners to pick his transaction, or again using his quantum device to gain an unfair mining advantage---to ensure his transaction is added to the block before Alice's. 

As long as Charly's quantum device can successfully calculate the private key given a public key comfortably with the blocktime of ten minutes, Charly would be able to consistently hijack Alice's, and anyone else's, entire UTXO.

The above assumes that Charly's device can break ECDSA in under 10 minutes. This is not unreasonable. ECDSA can be broken with a quantum circuit of depth \(O(m^2)\), where $m$ is the search space in bits\cite{cheung2008design}. Bitcoin uses a 256-bit ECDSA  key, therefore requiring approximately \(256^2 = 65,536\) logical gates. This would result in a quantum computer with a modest \emph{effective} $1$kHz clock-speed\footnote{By $1$kHz effective clock-speed we mean that the quantum computer can perform $1000$ logical operations per second. The physical clock-cycle required would depend on the number of error-correcting layers needed to achieve the required fault-tolerance.} taking approximately $66$ seconds to crack a vulnerable UTXO's ECDSA key.

Hence, unless accompanied by other major changes to the Bitcoin protocol, any process that aims to make Bitcoin quantum-secure by upgrading UTXOs to quantum-safe cryptosystems would have to be completed before the availability of scalable quantum computers.

\section{Summary of Main Results}

The main result of this paper is a strict, non-tight lower bound on the total \emph{cumulative} down-time required to perform an upgrade to the entire Bitcoin blockchain, replacing all current ECDSA-based UTXOs to post-quantum cryptography-based ones. This bound is  1827.96 hours, or 76.16 days---starting from the Bitcoin network as it stands today. If every single UTXO were upgraded to the Taproot Schnorr signature scheme\cite{BitcoinTaproot2021} \emph{before} the quantum-safety upgrade (current estimates of put the number of Taproot enabled UTXOs at less than $1\%$\cite{AdamISZ2024, Posch2023}), then this bound could be further lowered to $1306.80$ hours or $54.45$ days.

\begin{table}
    \centering
    \begin{tabular}{|c|c|c|c|c|}
            \hline
            \multirow{3}{*}{Bandwidth} & \multicolumn{4}{c|}{Lower Bound on Time Taken}\\
            \cline{2-5}
            & \multicolumn{2}{c|}{ECDSA-Based UTXOs} & \multicolumn{2}{c|}{Schnorr-Based UTXOs}\\
            \cline{2-5}
            & Hours & Days & Hours & Days \\
            \hline
            25\% & 7311.83 & 304.66 & 5227.18 & 217.80\\
            50\% & 3655.92 & 152.33 & 2613.59 & 108.90\\
            75\% & 2437.28 & 101.55 & 1742.39 & 72.60\\
            100\% & 1827.96 & 76.16 & 1306.80 & 54.45\\
            \hline
    \end{tabular}
    \caption{A lower bound on the downtime required to upgrade all vulnerable UTXOs to quantum-resistant, at varying levels of Bitcoin network bandwidth, shown in hours and days for both ECDSA-based and Schnorr-based UTXOs}
    \label{tab:bandwidths}
\end{table}

By \emph{cumulative} we mean to imply that the upgrade does not have to be done in a single, uninterrupted, 1827.96-hour session during which no regular Bitcoin transactions can be processed. Rather, the upgrade can be \emph{stretched} over a longer period. By doing this properly, rather than having strict downtime, Bitcoin would instead undergo \emph{throttling.} For instance,  the upgrade can be done in twice the time above---152.3 days---during which time Bitcoin's network would only be able to process half as many regular transactions per-time-period as usual. Table \ref{tab:bandwidths} presents several possible scenarios for this \emph{stretched} approach. 

We use the terms \emph{`strict'} and \emph{`non-tight'} in the usual mathematical sense. We ensured the bound is strict, regardless of any changes in technology, by setting \emph{all} transaction overhead costs to zero and assuming the most possibly (unrealistically) optimised packing of transactions into blocks. This makes the bound very loose. 

Despite their non-tight nature, the numbers presented in  Table \ref{tab:bandwidths} are the most central figures in this paper. This is because the bound does not rely on any assumptions---reasonable or otherwise---about technology or algorithms. So no advances in either will impact the bound. The only assumption the bounds in Table \ref{tab:bandwidths} make is that the Bitcoin protocol---besides upgrading its public-key cryptosystems to quantum-secure ones---remains the same. If the protocol were to be updated to, say, reduce the blocktime down from ten minutes, or increase the blocksize, these changes \emph{would} affect our bounds. In fact, one way to interpret our results is to conclude that Bitcoin \emph{will require} extensive changes, beyond \emph{just} the introduction of quantum-safe cryptosystems, in order to guarantee its future viability---\emph{i.e.} have it be safe against quantum attacks, and not undergo major throttling over large time-periods.

In the next section, we present the methodology and precise calculations we used to arrive to our results.

\section{Methodology \& Calculations}
\label{sec:txtime}

In this section, we derive the lower bounds on cumulative downtime (and/or throttling) required to make Bitcoin quantum-safe. These bounds were discussed in the previous section, and are summarised in Table  \ref{tab:bandwidths}.

\begin{table}
    \centering
    \begin{tabular}{|c|c|}
            \hline
            Field & Scale Factor\\
            \hline
            Version & 4\\
            Marker and Flag & 1\\
            Input & 4\\
            Output & 4\\
            Witness Data & 1\\
            Lock time & 4\\
            \hline
    \end{tabular}
    \caption{The conversion factor between bytes and weight units, obtained from the \textit{Mastering Bitcoin: Programming the Open Blockchain} textbook\cite{antonopoulos2017mastering} and modified to condense sub-fields into fields.}
    \label{tab:weights}
\end{table}
\begin{table}
    \centering
    \begin{tabular}{|c|c|c|c|}
            \hline
            Field & Size (Bytes) & Weight Units (WU) & Cumulative WU\\
            \hline
            Version & 4  & 16 & 16\\
            Marker and Flag & 2  & 2 & 18\\
            Input &  42 &  168 & 186\\
            Output & 44 &  176 & 362\\
            Witness Structure &  67 &  67 & 429\\
            Lock time & 4 & 16 & 445\\
            \hline
    \end{tabular}
    \caption{The structure of a transaction containing a single input and output, with sizes shown in bytes and weight units, obtained from the \textit{Mastering Bitcoin: Programming the Open Blockchain} textbook\cite{antonopoulos2017mastering} and modified to condense sub-fields into fields.}
    \label{tab:tx}
\end{table}

First, we must discuss any underlying assumptions behind our calculations. There is only one assumption made: aside from updating the digital-signature scheme currently used (ECDSA) to a post-quantum protocol, \emph{the Bitcoin network shall remain otherwise unchanged until the completion of the update.}
This is both a reasonable assumption to make, and a necessary one. 

Consider that any changes to the Bitcoin protocol must be approved by the majority of miners (typically 90\%), and the decentralised nature of the network makes such a consensus less than trivial. 
Hence, our goal is to demonstrate the cost, in downtime, of upgrading the Bitcoin network---as it currently stands---to quantum safe signature schemes. First, this showcases the costs involved in one of the---if not the---most, likely scenario. Second, it provides a useful baseline from which the impact/benefit of further updates to the network can be assessed.
All this said, the relevant parameters of the Bitcoin, at the time of writing (mid-2024) are as follows.

The available space in a given block on the Bitcoin network is defined in \textit{weight units (WU)}. Each block can contain a maximum of 4,000,000 WU, which itself can be allocated between block header information and individual transactions. The relationship between bytes and WU is not always a simple one-to-one mapping; rather, different weight unit conversion factors are applied to different fields within a transaction itself. \hyperref[tab:weights]{Table \ref{tab:weights}} shows the conversion factors applied to each major field of a transaction.

Three hundred and twenty (320) WU are reserved in each block for header information, and a transaction counter also requires up to 12 WU. However, to simplify the estimation of a lower bound, we have opted to omit this entirely; that is, we consider that all 4,000,000 WU can be used exclusively for upgrade transactions.

It is not a requirement that a single block must contain more than one transaction. In fact, considering that the Bitcoin community would want to optimise the upgrade process as much as possible (in order to reduce downtime), a solution would be to ensure that one block pertaining to UTXO upgrades contains a single transaction that is full of individual UTXO upgrades. This way, the metadata associated with an individual transaction is limited to just a single instance, allowing for many more UTXO upgrades per block.

Indeed, we can take this further by opting to omit metadata associated with a single transaction. On top of this, it can be argued that to maximise the number of inputs upgraded within a single block, the block itself must only reference a single output. This circumvents the need for an output associated with each input, effectively freeing up an additional 176 WU per input.

We acknowledge that some of these optimisations are impossible in practice; a block cannot exist without a header, for instance. As a direct result, the lower bound is effectively loosened. The point, however, is to show that even in an unrealistically optimal scenario, significant downtime is still required to ensure quantum safety.

\subsection{Crafting an Optimal Upgrade Block}
The key elements of an upgrade transaction are the input (i.e. the vulnerable UTXO that will be upgraded) and the output (i.e. the resulting quantum-secure UTXO after the upgrade process has completed). \hyperref[tab:tx]{Table \ref{tab:tx}} shows the various fields associated with a transaction that contains a single input and output. At a total size of 163 bytes, equating to 445 WU after applying each scale factor, this would allow for 8,988 transactions of this structure within a single block.

However, using the previously stated assumptions, we can populate a block with a \textit{single} transaction that contains several inputs and a single output, resulting in at most 17,020 UTXO upgrades per block.

\subsection{Calculating the Lower Bound}
\label{subsec:lb}
With approximately 186,676,874 UTXOs in circulation as of June 2024, and the  hard limit of 17,020 UTXO upgrades per block arrived at earlier, the rest of the calculation is pretty straightforward. 

The Bitcoin network aims to keep its \emph{blocktime} at ten minutes. This means that it takes approximately 10 minutes to add a new block to the blockchain. Let us assume that during the upgrade period, the entirety of every single block---\emph{i.e.} network bandwidth---is used for the upgrade. In this scenario, processing 186,676,874 UTXOs, at 17,020 UTXO upgrades per block, with one block added every 10 minutes, we arrive at  1,828 hours (76 days).
It is important to stress, once again, that this is a non-tight strict lower bound. Unless major changes are made to the Bitcoin network, actual processing times are likely to be considerably higher. 

It is unlikely that the Bitcoin network would survive, as a viable cryptocurrency, seventy-six days of uninterrupted downtime. A more feasible approach is to use only a percentage of each block for upgrade transactions, in essence \emph{throttling} the Bitcoin network during the upgrade process. 
 \hyperref[tab:bandwidths]{Table \ref{tab:bandwidths}} shows lower bounds on the total upgrade process time with varying levels of network bandwidth

Finally, let us consider the Taproot upgrade, which can impact these transaction times. Implemented in 2021\cite{BitcoinTaproot2021}, Taproot offers many benefits, such as increased security and scalability. Of major importance to us is the introduction of Schnorr signatures. Transactions using Taproot have a smaller footprint on the blockchain. This directly impacts our lower bound above.

However, there are two major issues with Taproot and Schnorr signatures in particular. First, 
 the Schnorr signature scheme, just like ECDSA, is based on the discrete logarithm problem---and as a result, is vulnerable to quantum attacks.

 The second issue is that for the space-savings mentioned above to happen, the input (sender) in the transaction has to have a Taproot enabled Schnorr signature. Currently, Taproot is used by approximately  0.1\% - 1\% of all transactions\cite{AdamISZ2024, Posch2023}. In other words, if the upgrade to quantum-safety is to benefit significantly from the Taproot upgrade, then most if not all current UTXOs have to be `upgraded' to Taproot. 
 
 This upgrade-to-Taproot might seem redundant with little to no advantage, then. However, upgrading all UTXOs to Taproot is a process that can begin \emph{now}---and in a sense, it has already begun, even if Taproot adoption has been significantly slowing down since its introduction\cite{kerr2023taproot}---without having to wait for a quantum signature scheme for Bitcoin to be adopted. This advantage may come into play, as we discuss further below. 
 
 Hence, it is worth calculating the precise benefit of using Taproot for the quantum-safety upgrade. In order to calculate the time taken for the upgrade process while using Taproot, the same methodology is used as when calculating for the Segwit transactions. We assume that as many UTXOs are compiled into one single input as possible, directed at a single output. Additionally, as Schnorr signatures are capable of key aggregation, only a single signature and witness would be required for all the inputs. We thus calculate that 23,807 UTXOs can fit into a single block, resulting in 7841 blocks of upgrade transactions. This equates to $1306.80$ hours or $54.45$ days.  Again, this is under the assumption that all current UTXOs have been upgraded to the Taproot Schnorr signature scheme. If only a portion of current UTXOs have been upgraded, the real lower bound will be a linear combination of this number, and the previously calculated one, weighted by the proportion of Schnorr signature-enabled UTXOs. See \hyperref[tab:bandwidths]{Table \ref{tab:bandwidths}}.

\section{Discussion}
\label{sec:disc}

Next, we consider the implications of previous results within the greater Bitcoin, cryptocurrency, and cybersecurity contexts.

First, it is worth repeating that the 76-day downtime figure is merely a lower bound. For this paper, we focused on deriving a strict, if non-tight bound for several reasons---the most central being the unconditional correctness of the resulting bound. An even casual reading of the previous section should reveal many areas where we left room for the bound to be tightened in future work. Beyond that, however, is that any actual real-world implementation of the upgrade process is very unlikely to be anywhere close to what a lower bound that assumes optimal processes throughout would suggest.

Second, even if we were to completely disregard the non-tight bound---based on unrealistically optimal assumed procedures---nature of the 76-day figure, the fact still remains that 76 days is an unreasonably long amount of time  for the Bitcoin network to be \emph{``brought down for maintenance.''} Hence, it is unavoidable that the upgrade procedure will have to be interwoven with regular Bitcoin transactions. 

The obvious approach to this would be to designate every $k'th$ block to quantum-safe upgrade transactions\footnote{The alternative is to allocate a $k'th$-fraction of each block for upgrade transactions. Which of these approaches is more optimal is an open question, whose answer would depend on the particulars of the packing algorithms used, and real-world data. Regardless, it is highly unlikely for one to provide a sizeable advantage over the other.}. However, the transaction bandwidth of Bitcoin is partitioned, this brings us to the third and fourth concerns: Bitcoin's service quality, and the upgrade deadline.

 At the time of writing, the average time for a transaction in the  pool to be added to a block is 576 minutes. This time has been as high as 12,404 minutes (206 hours) as recently as October 2023\cite{coinmarketcap2022bitcoin}. This has been a persistent point of criticism for the Bitcoin network  throughout its lifetime\cite{khan2021challenges, jani2020scalability}; and it has only become worse as  the technology has increased in use and popularity. 

 Obviously, allocating a $k'th$ portion of Bitcoin's transaction bandwidth to upgrade transactions will result in an unavoidable $1/k$ increase in transaction times. On a cryptocurrency which is already known for its slow transaction speed, the impetus would be to make $k$ as small as viably possible. This, however, runs counter to our next area of concern.

 The existence of the Just-in-Time quantum attack detailed in this paper means that, to ensure the integrity of the Bitcoin network, the entire upgrade process has to be completed before quantum computers capable of breaking ECDSA-256 are brought into production. A recent survey of quantum computer experts concluded that quantum computers capable of breaking ECDSA and other public-key cryptosystems can be expected, with roughly $31\%$ probability, within the next decade\cite{mosca2023quantum}.
  All this means that for the upgrade to be completed securely, with as small an impact on Bitcoin service-quality as possible, the upgrade needs to be commenced as soon as possible. This brings us to our next concern.

 The fifth issue is that before any upgrade process can be commenced, a ($90\%$) consensus among Bitcoin miners has to be achieved over the particular details of the upgrade. Historically, considerable changes to the Bitcoin network have been met with high resistance. A notable example of this was the SegWit upgrade in 2017\cite{narayanan2017bitcoin}. This upgrade caused the Bitcoin community to disagree to such an extent that several hard forks were performed, creating Bitcoin Cash\cite{bitcoincash} and Bitcoin Gold\cite{bitcoingold}. The high agreement threshold that needs to be surmounted in order to trigger a Bitcoin upgrade protocol means that consensus is often slow to achieve---even in the best of cases where there are little to no controversial decisions to be made. This particular upgrade that we're discussing here, however, is likely to stir up quite a bit of controversy---if for nothing else, the choice of ECDSA-256 replacement. Which brings us to our next major point of concern.

Among all the other details and parameters of the upgraded Bitcoin protocol, and its upgrade process, that need to be decided upon, a central and crucial one is the post-quantum signature scheme that is to replace ECDSA-256 in Bitcoin's UTXOs. In short, there is currently \emph{no} protocol that is likely to appeal to the majority of the Bitcoin user-base.

There are currently \emph{three} post-quantum signature schemes (and one encryption scheme) that have been approved as a standard by  the US National Institute of Standards and Technology (NIST) following its 2017 call for proposals\cite{NIST2016} for quantum-secure encryption and signature schemes to replace the current public-key cryptosystems. The three standardized signature schemes are  Crystals-Dilithium\cite{Ducas2018},  Falcon\cite{Fouque2018}, and SPHINCS+\cite{Bernstein2019}. The standards for Kyber\cite{FIPS203}, Dilithium\cite{FIPS204} and SPHINCS+\cite{FIPS205} have since been released, with Falcon's set to follow. All three schemes allow for a parametrised signature length to be chosen as a tradeoff between performance and security.  Table \ref{table:signatures} shows the signature size required, for each of the three standardised post-quantum signature schemes, to achieve the same level of security as ECDSA-256. As can be surmised from the table, even the most size-efficient post-quantum signature scheme---FALCON---produces signatures that are ten times longer than ECDSA-256.

\begin{table}[h]
\centering
\begin{tabular}{| m{5cm} | m{5cm} |}
  \hline
  \textbf{Digital Signature Scheme} & \textbf{Signature Size (bits)} \\
  \hline
  CRYSTALS-Dilithium & 19360 \\
  \hline
  FALCON & 5328 \\
  \hline
  SPHINCS+ & 62848 \\
  \hline
  ECDSA  & 512 \\
  \hline
\end{tabular}
\caption{NIST Post-Quantum Cryptography Standardization Digital Signature Schemes and Their Signature Sizes}
\label{table:signatures}
\end{table}

This increase in size does not necessarily impact the \emph{upgrade} process, since these transactions all use \emph{current} ECDSA signatures as inputs, and use \emph{hashes} of the post-quantum signatures as outputs---and these hashes do not necessarily have to change in size.

However, once the upgrade is underway, any new transactions that use the \emph{upgraded} quantum-safe UTXOs as \emph{inputs} will necessarily use these much longer quantum-safe signatures. This will lower the number of transactions that will fit in a block; necessarily slowing down the Bitcoin transaction time---all the way to a possible \emph{permanent} 10x slowdown once the upgrade is completed.

Given the current state of post-quantum signature schemes, it is then tempting to wait for newer, more efficient, schemes to come out in the future. It is important to note here that  NIST made an additional call for post-quantum digital signature proposals, which closed in June 2023\cite{NIST2022}. Additional NIST standards are expected to arrive by the end of 2025\cite{nist_pqc_2023}. However, even then, these new signature schemes may not provide smaller signatures: the aim of this additional call is to diversify the post-quantum signature standards, since three of the four aforementioned schemes are based on structured lattices. The main driving force behind this diversification is security, not necessarily efficiency. Regardless, any attempt to wait for improved quantum-safe signature technology is necessarily going to make the upgrade process more damaging to the Bitcoin quality-of-service.

\section{Conclusion}
\label{sec:conc}
In this work, we have highlighted that the very real threat of quantum attack on the Bitcoin network demands immediate action. Through crafting a theoretically optimal upgrade block, we have proposed a 76-day (54-day with Schnorr-based UTXOs considered) non-tight lower bound to the amount of time required to fully upgrade all UTXOs. This is under the assumption that 100\% network bandwidth is reserved for upgrade transactions; in reality, a more conservative bandwidth allowance (of, say, 25\%) would result in a minimum of approximately 305 days to perform the complete upgrade.

Such a bandwidth allowance would effectively result in a reduction of Bitcoin efficiency by 25\%, leading to an increase in transaction fees, slower transaction confirmation times and a decrease in market confidence. However, the alternative to this; that is, a reality in which no upgrade to post-quantum resistance is performed, would be devastating. Through the use of a quantum JIT attack, a threat actor with sufficient quantum computing capability could, with a modestly powered quantum device, attack arbitrary outgoing transactions and hijack the transacted Bitcoin. In such a scenario, Bitcoin (indeed, any cryptocurrency relying on a cryptosystem that is broken by advances in quantum computing) as we know it would cease to exist. 

These calculations do not account for the considerable amount of time that would be required for research, discussion, decision-making, and implementation. Furthermore, other areas of the blockchain protocol that are vulnerable\cite{Bard2022} are not considered. This could potentially be one of the most divisive upgrades in the history of Bitcoin, and the process of beginning to upgrade could take a considerable period of time. 

It is important to note that whilst we cannot say with absolute certainty as to when a quantum computer capable of attacking the Bitcoin network is developed, we can be confident that such a device will inevitably exist. Once this date arrives, any existing infrastructure (whether relating to Bitcoin or any other application relying on problems that are solvable in polynomial time by quantum algorithms) can be considered vulnerable. In the case of the Bitcoin network, it is imperative that this upgrade is completed proactively rather than reactively to ensure the safety of all UTXOs on the network.

\bibliographystyle{splncs04}
\bibliography{bibliography}

\end{document}